\documentclass[a4paper,11pt]{article}
\usepackage{jinstpub}
\usepackage{lineno}
\usepackage{gensymb}
\usepackage{subcaption}
\usepackage{ifthen}
\newcommand{\cmsorcid}[1]{%
  \ifthenelse{\equal{#1}{}}%
  {}% Do nothing if empty
  {%
    \href{https://orcid.org/#1}{\raisebox{0.5ex}{\includegraphics[width=0.7em]{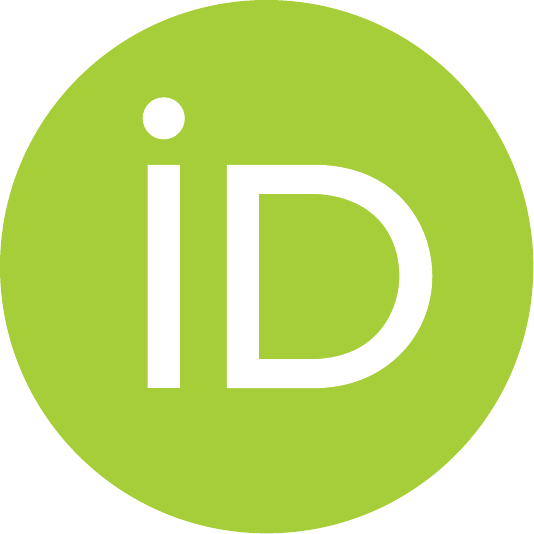}}}%
  }%
}

\title{\boldmath Annealing behaviour of charge collection of neutron irradiated diodes from 8-inch p-type silicon wafers}

\author[a,b,1]{Oliwia~Agnieszka~Ka\l{}uzi\'{n}ska\cmsorcid{0009-0001-9010-8028}\note{Corresponding author.}}
\author[a]{Leena~Diehl\cmsorcid{0000-0002-7962-0661}}
\author[a]{Eva~Sicking\cmsorcid{0000-0002-4025-2566}}
\author[a,c]{Marie~Christin~M\"uhlnikel\cmsorcid{0009-0007-6437-3542}}
\author[a,d]{Pedro~Gon\c{c}alo~Dias~de~Almeida\cmsorcid{0000-0001-8787-9529}}
\author[a,e]{Jan~Kieseler\cmsorcid{0000-0003-1644-7678}}
\author[a,f]{Matthias~Kettner\cmsorcid{0009-0002-4586-5350}}
\author[a]{David~Walter\cmsorcid{0000-0001-8584-9705}}
\author[a]{Matteo~M.~Defranchis\cmsorcid{0000-0001-9573-3714}}

\affiliation[a]{Experimental Physics department, CERN,\\Espl. des Particules 1, Geneva, Switzerland}
\affiliation[b]{ETIT, Karlsruhe Institute of Technology,\\Engelbert-Arnold-Straße 4, Karlsruhe, Germany}
\affiliation[c]{Kirchhoff Institute for Physics, Heidelberg University,\\Im Neuenheimer Feld 227, Heidelberg, Germany}
\affiliation[d]{Instituto de Física de Cantabria (CSIC-UC),\\Av. de los Castros, Santander, Spain}
\affiliation[e]{EPT, Karlsruhe Institute of Technology,\\Wolfgang-Gaede-Str. 1, Karlsruhe, Germany}
\affiliation[f]{Institute of High Energy Physics of the Austrian Academy of Sciences,\\Dominikanerbastei 16, A-1010, Vienna, Austria}

\emailAdd{oliwia.agnieszka.kaluzinska@cern.ch} 

\abstract{To face the higher levels of radiation due to the 10-fold increase in integrated luminosity during the High-Luminosity LHC, the CMS detector will replace the current  Calorimeter Endcap (CE) using the High-Granularity Calorimeter (HGCAL) concept.
The electromagnetic section as well as the high-radiation regions of the hadronic section of the CE will be equipped with silicon pad sensors, covering a total area of 620 $\rm m^2$.
Fluences up to $\rm1.0\cdot10^{16}~n_{eq}/cm^{2}$ and doses up to 2 MGy are expected considering an integrated luminosity of 3 $\rm ab^{-1}$.
The whole CE will normally operate at -35\degree C in order to mitigate the effects of radiation damage.

The silicon sensors are processed on 8-inch p-type wafers with an active thickness of 300~\textmu m, 200~\textmu m and 120~\textmu m and cut into hexagonal shapes for optimal use of the wafer area and tiling.
With each main sensor several small sized test structures (e.g pad diodes) are hosted on the wafers, used for quality assurance and radiation hardness tests.
In order to investigate the radiation-induced bulk damage, 28 diodes have been irradiated with reactor neutrons at the TRIGA reactor in JSI (Jožef Stefan Institute, Ljubljana) to 13 fluences between $\rm6.5\cdot10^{14}~n_{eq}/cm^{2}$ and $\rm1.5\cdot10^{16}~n_{eq}/cm^{2}$.

The charge collection of the irradiated silicon diodes was determined through transient current technique (TCT) measurements.
The study focuses on the isothermal annealing behaviour of the bulk material at 60\degree C.
The results have been used to extend the usage of thicker silicon sensors in regions expecting higher fluences and are being used to estimate the expected annealing effects of the silicon sensors during year-end technical stops and long HL-LHC shutdowns currently foreseen with a temperature around~0\degree C.}

\keywords{Calorimeters, Radiation damage to detector materials (solid state), Radiation-hard detectors, Solid state detectors}

\arxivnumber{} % Only if you have one

\begin{document}

%%%%%%%%%%%%%%%
\newcommand{\neqcm}{10\textsuperscript{15} n\textsubscript{eq}/cm\textsuperscript{2}}
%%%%%%%%%%%%%%%
\maketitle
\flushbottom

\section{Introduction}
\label{sec:intro}

The High-Luminosity Large Hadron Collider (HL-LHC) upgrade aims at enhancing the performance of the LHC, the most powerful particle accelerator in the world, to boost the potential for scientific discoveries starting from 2030.
The project includes increasing the luminosity (and thus the collision rate) by a factor of five beyond the original LHC design specifications.
The integrated luminosity will be increased by a factor of ten.
This poses significant challenges in terms of radiation tolerance and event pileup for the detectors~\cite{Apollinari:2284929, HGCAL-TDR}.

As part of the HL-LHC upgrade within the CMS Experiment~\cite{CMS_det_paper}, the current Calorimeter Endcap (CE) will be replaced, using the novel High Granularity Calorimeter (HGCAL) concept.
It will consist of 47 sampling layers interleaved with Cu, CuW, Pb and stainless steel absorber plates and will include nearly six million readout channels.

Silicon sensors were selected as active material for the majority of the CE upgrade due to their compactness, rapid signal formation, and adequate radiation hardness.
At the HL-LHC, these silicon sensors will be subjected to hadron fluences ranging from approximately $\rm2.0\cdot10^{14}$ 1-MeV neutron-equivalents per square centimeter $\rm (n_{eq}/cm^{2})$ to $\rm1.0\cdot10^{16} n_{eq}/cm^{2}$ after an integrated luminosity of 3000 $\rm fb^{-1}$.
Around 620 $\rm m^2$ of silicon sensors will cover the entire electromagnetic (CE-E) section and the high-radiation region of the hadronic (CE-H) section of the calorimeter.
The silicon sensors are fabricated on 8-inch wafers and diced to form a hexagonal shape for efficient use of the wafer area and tiling~\cite{CMSHGCAL:2022mtx, HGCAL-TDR}.

In this paper, we report about three measurement campaigns investigating isothermal annealing behaviour of the charge collection capability of the silicon bulk material using dedicated test structure 5x5 $\rm mm^2$ diodes from the edge region of the wafer, irradiated with reactor neutrons up to $\rm1.5\cdot10^{16} n_{eq}/cm^{2}$.
As the CMS detector region to be occupied by CE will be irradiated  predominantly by neutrons~\cite{HGCAL-TDR}, these particles were the main focus of the irradiation.
The campaigns were performed in 2021 and 2023 in two different measurement setups.
Section~\ref{sec:sensors} gives an overview and detailed information about the samples used in each campaign.
The measurement setups and techniques are described in section~\ref{sec:setups}.
An insight into the data analysis procedure is presented in section~\ref{sec:analysis}.
The discussion on the systematic uncertainties is provided in section~\ref{sec:uncertainties}.
The results and interpretation of the obtained data can be found in section~\ref{sec:results}.
Finally, the summary and conclusions are given in section~\ref{sec:discussion}.
\section{Samples}
\label{sec:sensors}

\begin{figure}[h]
    \centering
    \includegraphics[width=0.59\linewidth]{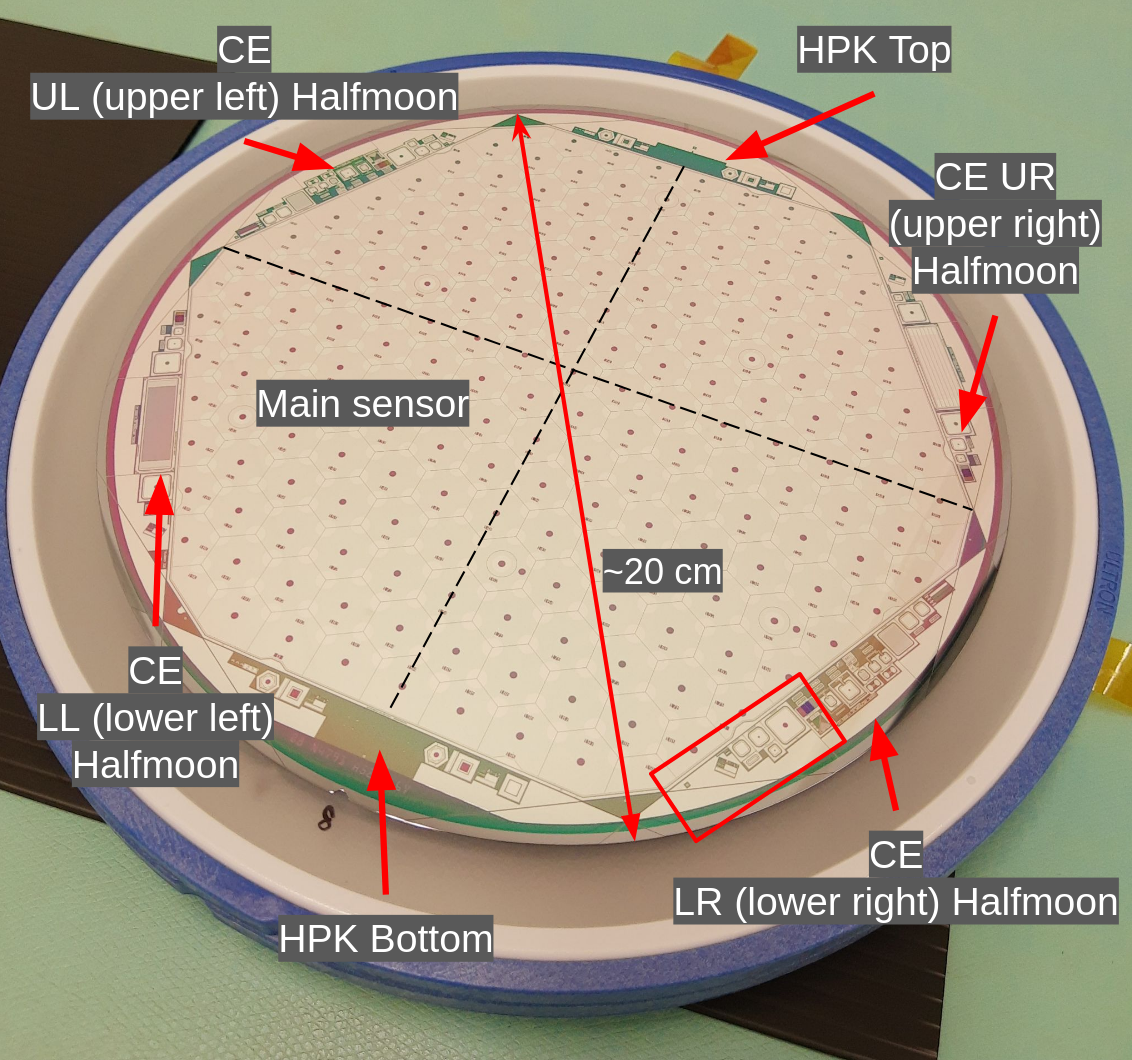}
    \caption{CE full sensor on 8-inch circular wafer including test structures. The silicon crystal orientation is presented as black dashed lines. The example halfmoon section containing the diode used in this publication is marked in red frame (the diode used has cross-sectional area 5x5 $\rm mm^2$). Reproduced from \cite{Kaluzinska:2815709}.}
    \label{fig:HGCAL_full_wafer}
\end{figure}

The silicon sensors to be used in the CMS Calorimeter Endcap upgrade consist of DC-coupled, planar, high resistivity (>3 k\ohm cm), p-type hexagonal silicon sensors with a crystal orientation of <100> produced on 8-inch circular wafers by Hamamatsu Photonics K.K\footnote{\url{https://www.hamamatsu.com/eu/en.html}} (shown in figure~\ref{fig:HGCAL_full_wafer})~\cite{HGCAL-TDR}.
Sensors are produced in three different active thicknesses and two material types.
300~\textmu m and 200~\textmu m sensors are produced in the float zone (FZ) process, whereas 120 \textmu m sensors are produced in the epitaxial (EPI) process on top of a handling wafer of $\sim$180 \textmu m thickness.
The remaining space left around the hexagonal main sensor (called "halfmoons") is used for the fabrication of small-sized test structures, which have an identical production process as the main sensor.
The test structures include, among others, single diodes which can be used to investigate the bulk radiation hardness of the sensors.
For the measurements presented here, square diodes with a side length of 5~mm are chosen.
The (main) n+ pad implant is surrounded by a guard ring.

In the following, results from three independent charge collection measurement campaigns are presented, performed in two different setups referred to as "TCT+ setup" and "Particulars setup"\footnote{\url{https://particulars.si/index.php}}. 
The setups are described in detail in section~\ref{sec:setups}.
In the TCT+ setup, two measurement campaigns were performed - one in 2021 with CE version-1 prototype samples, irradiated to intermediate fluences ($\rm6.5\cdot10^{14} n_{eq}/cm^{2}$ to $\rm1.0\cdot10 ^{16} n_{eq}/cm^{2}$) and one in early 2023 with CE version-2 prototype samples (and with additional version-1 prototype 120 \textmu m samples), irradiated to higher fluences ($\rm1.5\cdot10^{15} n_{eq}/cm^{2}$ to $\rm1.4\cdot10 ^{16} n_{eq}/cm^{2} $).
The third campaign was performed in the Particulars setup in 2023, where CE pre-series samples, irradiated to higher fluences ($\rm2.0\cdot10^{15} n_{eq}/cm^{2}$ to $\rm1.5\cdot10 ^{16} n_{eq}/cm^{2} $), were measured.
Version-1, version-2 prototypes and pre-series sensors represent different iterations of sensors produced for CE.
The iterations differ in terms of parameters regarding the oxidation process.
Typical variations in the bulk production process are known to be independent of the manufacturer's oxidation process.
The bulk properties of the silicon material are defined in the procurement documents and are expected to remain consistent across production iterations.
All wafers are certified to meet the technical requirements specified in the invitation to tender.
As a result, consistent performance in charge collection measurements is expected.
In each campaign, isothermal annealing studies were performed.
Detailed information about the samples and annealing steps for each campaign is listed in tables~\ref{tab:2021_SSD}, \ref{tab:2023_SSD} and \ref{tab:2023_Particulars}, respectively.
All diodes were irradiated with reactor neutrons at the Training, Research, Isotopes, General Atomics (TRIGA) reactor located at the Jožef Stefan Institute, Ljubljana, Slovenia~\cite{SNOJ2012483, AMBROZIC2017140}.
In addition to the irradiated samples, each campaign included a set of unirradiated samples - one for each thickness - to serve as reference sensors.

\begin{table}
    \centering
    \caption{Data set overview for 2021 TCT+ campaign. `LR' refers to the lower right
halfmoon of the wafer (ref.\ fig.\ \ref{fig:HGCAL_full_wafer}). The ``Est. initial 60$^{\circ}$C offset'' is explained in the text. For the last annealing step, the samples were annealed in two batches with different times as indicated in the last row.}
    \begin{tabular}{r|rrrr}
        \textbf{Sensor ID} & \textbf{Active} & \textbf{Fluence} & \textbf{Irradiation} & \textbf{Est. initial 60\degree C}\\
        & \textbf{thickness [\textmu m]} & \textbf{[\neqcm]} & \textbf{time [min]} & \textbf{ offset [min]} \\
        \hline
        1002\_LR & 300 & 0.65 & 7.4 & 1.0 $\pm^{0.8}_{0.4}$\\
        1003\_LR & 300 & 1.00 & 10.8 & 1.6 $\pm^{1.3}_{0.7}$\\
        1102\_LR & 300 & 1.50 & 16.2 & 2.4 $\pm^{2.1}_{1.1}$\\
        2002\_LR & 200 & 1.00 & 10.8 & 1.6 $\pm^{1.3}_{0.7}$\\
        2003\_LR & 200 & 1.50 & 16.2 & 2.4 $\pm^{2.1}_{1.1}$\\
        2102\_LR & 200 & 2.50 & 27.5 & 4.0 $\pm^{3.8}_{1.9}$\\
        3008\_LR & 120 & 1.50 & 16.2 & 2.4 $\pm^{2.1}_{1.1}$\\
        3007\_LR & 120 & 2.50 & 27.5 & 4.0 $\pm^{3.8}_{1.9}$\\
        3003\_LR & 120 & 10.00 & 108.0 & 14.2 $\pm^{14.2}_{7.2}$\\
    \hline
    Annealing steps [min] & \multicolumn{4}{|c}{10, 30, 90, 120 (all samples),} \\
     & \multicolumn{4}{|c}{260 (1003\_LR, 2002\_LR, 3008\_LR, 3003\_LR),} \\
     & \multicolumn{4}{|c}{300 (1002\_LR, 1102\_LR, 2003\_LR, 2102\_LR, 3007\_LR)} \\
    \end{tabular}
    \label{tab:2021_SSD}
\end{table}

\begin{table}
    \centering
    \caption{Data set overview for 2023 TCT+ campaign. `LR' refers to the lower right, `LL' refers to the lower left, `UR' to upper right and `UL' to upper left halfmoon of the wafer (ref.\ fig.\ \ref{fig:HGCAL_full_wafer}). The ``Est. initial 60$^{\circ}$C offset'' is explained in the text.}
    \begin{tabular}{r|rrrr}
        \textbf{Sensor ID} & \textbf{Active} & \textbf{Fluence} & \textbf{Irradiation} & \textbf{Est. initial 60\degree C}\\
        & \textbf{thickness [\textmu m]} & \textbf{[\neqcm]} & \textbf{time [min]} & \textbf{ offset [min]} \\
        \hline
        N4791\_11\_LR & 300 & 1.50 & 16.2 & 2.4 $\pm^{2.1}_{1.1}$\\
        N4791\_12\_LL & 300 & 2.00 & 21.8 & 3.2 $\pm^{3.0}_{1.5}$\\
        N4791\_13\_LR & 300 & 3.00 & 32.6 & 4.7 $\pm^{4.4}_{2.3}$\\
        N4792\_10\_LR & 200 & 2.00 & 21.8 & 3.2 $\pm^{3.0}_{1.5}$\\
        N4792\_11\_LL & 200 & 4.00 & 43.3 & 6.1 $\pm^{5.9}_{3.0}$\\
        N4792\_12\_LR & 200 & 5.50 & 59.5 & 8.1 $\pm^{8.0}_{4.1}$\\
        N4789\_10\_LL & 120 & 5.50 & 59.5 & 8.1 $\pm^{8.0}_{4.1}$\\
        N4789\_12\_UL & 120 & 10.00 & 108.0 & 14.2 $\pm^{14.2}_{7.2}$\\
        N4789\_13\_UR & 120 & 14.00 & 151.2 & 19.6 $\pm^{19.8}_{10.0}$\\
        3008\_UR & 120 & 1.50 & 16.2 & 2.4 $\pm^{2.1}_{1.1}$\\
        3007\_UR & 120 & 2.50 & 27.5 & 4.0 $\pm^{3.8}_{1.9}$\\
        3003\_UR & 120 & 10.00 & 108.0 & 14.2 $\pm^{14.2}_{7.2}$\\
    \hline
    Annealing steps [min] & \multicolumn{4}{|c}{0, 90} \\
    \end{tabular}
    \label{tab:2023_SSD}
\end{table}

\begin{table}
    \centering
    \caption{Data set overview for 2023 Particulars campaign. `LR' refers to the lower right, `LL' refers to the lower left and `UL' to upper left halfmoon of the wafer (ref.\ fig.\ \ref{fig:HGCAL_full_wafer}). The ``Est. initial 60$^{\circ}$C offset'' is explained in the text.}
    \begin{tabular}{r|rrrr}
        \textbf{Sensor ID} & \textbf{Active } & \textbf{Fluence} & \textbf{Irradiation} & \textbf{Est. initial 60\degree C}\\
        & \textbf{thickness [\textmu m]} & \textbf{[\neqcm]} & \textbf{time [min]} & \textbf{ offset [min]} \\
        \hline
        N8738\_1 LL1 & 300 & 2.00 & 21.8 & 3.2 $\pm^{3.0}_{1.5}$\\
        N8738\_2 LR & 300 & 4.00 & 43.3 & 6.1 $\pm^{5.9}_{3.0}$\\
        N8740\_1 LL1 & 200 & 4.00 & 43.3 & 6.1 $\pm^{5.9}_{3.0}$\\
        N8740\_2 LL1 & 200 & 6.00 & 64.9 & 8.8 $\pm^{8.7}_{4.4}$\\
        N8740\_3 LL2 & 200 & 8.00 & 86.8 & 11.6 $\pm^{11.5}_{5.8}$\\
        N8737\_3 UL & 120 & 6.00 & 64.9 & 8.8 $\pm^{8.7}_{4.4}$\\
        N8737\_2 LL1 & 120 & 15.00 & 161.8 & 20.9 $\pm^{21.1}_{10.7}$\\
    \hline
    Annealing steps [min] & \multicolumn{4}{|c}{0, 30, 55, 85, 120, 150, 200, 275, 365, 510} \\
    \end{tabular}
    \label{tab:2023_Particulars}
\end{table}

During the irradiation, the diodes are exposed to temperatures reaching up to ($50 \pm 5$)\degree C.
This is taken into account as in-reactor annealing when determining the total annealing times and its uncertainty for the individual samples.
The temperature during the irradiation is assumed to increase from the minimum of 25\degree C to the maximum temperature after 30 min where it stabilizes.
The temperature increase and decrease is assumed to follow exponential functions, with distinct time constants for each phase~\cite{Kieseler_2023}. Based on this information, the irradiation time, listed in tables~\ref{tab:2021_SSD}, \ref{tab:2023_SSD} and \ref{tab:2023_Particulars}, is converted to an equivalent annealing time at 60\degree C (using the so-called Hamburg model derived in~\cite{Moll:1999kv, Moll:2018fol}) and included in the tables as \textit{estimated initial 60\degree C offset}.
The estimated annealing times have an uncertainty which is displayed in the figures as error bars.

\begin{figure}
    \centering
    \begin{subfigure}{0.47\textwidth}
    \centering
        \includegraphics[width=0.85\textwidth]{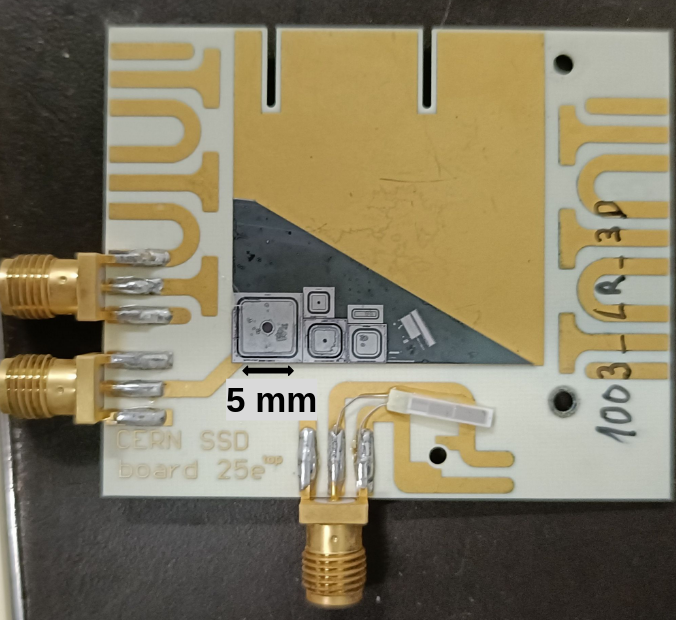}
        \caption{Large custom designed 2-layer PCB layout. The full test structure part fits onto the PCB.}
        \label{fig:PCB25}
    \end{subfigure}
    \hfill
    \begin{subfigure}{0.47\textwidth}
    \centering
         \includegraphics[width=0.95\textwidth]{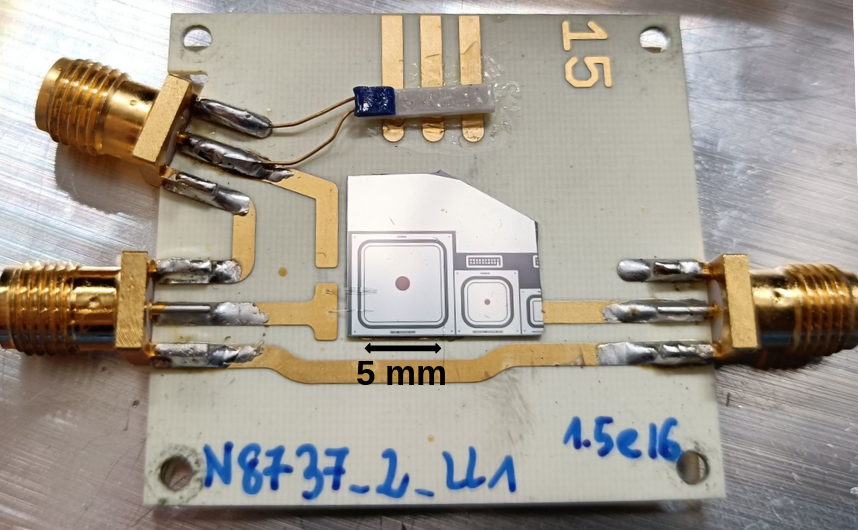}
        \caption{Small custom designed 2-layer PCB layout. The test structure sample has to be diced to fit onto the PCB.}
         \label{fig:PCB15}
    \end{subfigure}
    \caption{Two PCB layouts used in this study.}
    \label{fig:PCB_layouts}
\end{figure}

After the irradiation, samples are mounted on a printed circuit board (PCB) with a conductive silver paint.
The diode is wire-bonded to a SubMiniature version A (SMA) connector enabling appropriate connection to the measurement circuit through which it is grounded.
The diode's guard ring is grounded.
The high voltage is applied to the diode's backside.
In order to control the temperature of the diode, in the following referred to as device under test (DUT), during the measurement, a PT1000 resistor is glued close to the DUT.
In the 2021 TCT+ campaign, a large custom designed 2-layer PCB layout, shown in figure~\ref{fig:PCB25}, was used.
Such a large PCB size was used to avoid further dicing the pre-diced triangles of the halfmoons containing the DUT.
However, it was found non-optimal for homogeneous cooling of the PCB area, especially below the DUT, as it is not fully covered by the copper holder that provides cooling (ref.\ section~\ref{sec:setups}).
For the next two campaigns (both 2023 campaigns), a small custom designed 2-layer PCB layout (shown in figure~\ref{fig:PCB15}) was chosen to improve homogeneity of the temperature across the DUT.
To fit the size of smaller PCB, the silicon triangles had to be diced, which was performed using a diamond scribe.
This posed the risk of damaging the diode intended for testing during the cutting especially since the silicon crystal orientation and the diode orientation differ by 30 degrees, as can be seen in figure~\ref{fig:HGCAL_full_wafer}.

Between measurements, the samples were annealed in several steps.
The annealing is performed in a dedicated oven preheated to a temperature of 60\degree C.
For the annealing, the samples are placed on preheated copper blocks for good thermal contact.
The temperature was monitored every second using the PT1000 resistor attached to one of the PCBs.
The estimated initial offset is added to the additional annealing steps resulting in the total annealing time.
The additional annealing steps are listed in the bottom row of tables~\ref{tab:2021_SSD}, \ref{tab:2023_SSD} and \ref{tab:2023_Particulars}.
Except during annealing or measurement, the samples are stored in a freezer (at -20\degree C) to avoid excess annealing.
\section{Measurement setups}
\label{sec:setups}

The ability to collect charge by the silicon sensors can be measured using the Transient Current Technique (TCT)~\cite{Kramberger:1390490}.
In this method, a laser pulse generates electron-hole pairs in the silicon bulk.
As an external bias voltage is applied to the silicon diode, the generated charge carriers drift towards the corresponding electrodes.
The induced current is recorded as a function of time. 

\begin{figure}[h]
    \centering
    \includegraphics[width=0.55\textwidth]{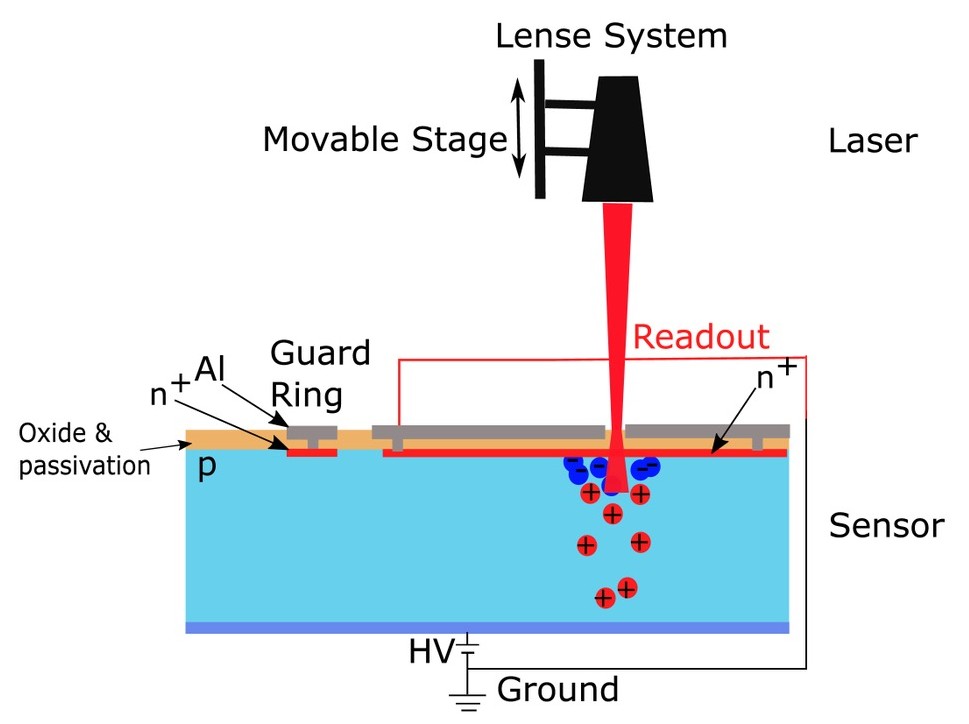}
    \caption{Schematic view of top-TCT measurement (adapted from~\cite{dissLeena}).}
    \label{fig:TCT_principle}
\end{figure}

For all measurements reported in this publication, the DUTs are reverse-biased, with high voltage applied to the back side bulk contact.
The readout is connected to an oscilloscope via an amplifier.
Laser light of 1064~nm wavelength (infra-red) is injected from the front side of the DUT (IR-top TCT) directed at a 1mm-diameter hole in centre of the metal contact layer of the diode.
In the TCT+ setup, the laser had a repetition frequency of 200 Hz and a duty cycle of 0.002\%, whereas, in the Particulars setup, the repetition frequency was 1 kHz with a duty cycle of 0.0002\%.
The schematic principle of the top-TCT measurement is shown in figure~\ref{fig:TCT_principle}.
In the 2021 TCT+ campaign, a laser intensity equivalent to 33 MIPs (minimum ionizing particles) was used.
For the 2023 TCT+ and Particulars campaigns, the intensity was increased to 40 MIPs, which is still within the linear region of the laser intensity where plasma effects are no concern, in order to enhance the signal for highly irradiated diodes~\cite{Plasma:9954488}.
In the previous studies (\cite{HGCAL-TDR, dissEsteban}), it was shown that an IR light signal equivalent to multiple MIPs gives comparable results to single-MIP measurements using a radioactive source.
The measurements are conducted with the DUT cooled down to -20\degree C in order to reduce the leakage current of the irradiated sensors.
The difference of the measurement temperature compared to the -35\degree C operation temperature has no influence on the charge collection efficiency (CCE, ref. section~\ref{sec:results}), while for the leakage current the current can easily be scaled.
This was cross-checked by measurements using modules assembled with irradiated sensors.

\begin{figure}[h]
    \centering
    \includegraphics[width=1\textwidth]{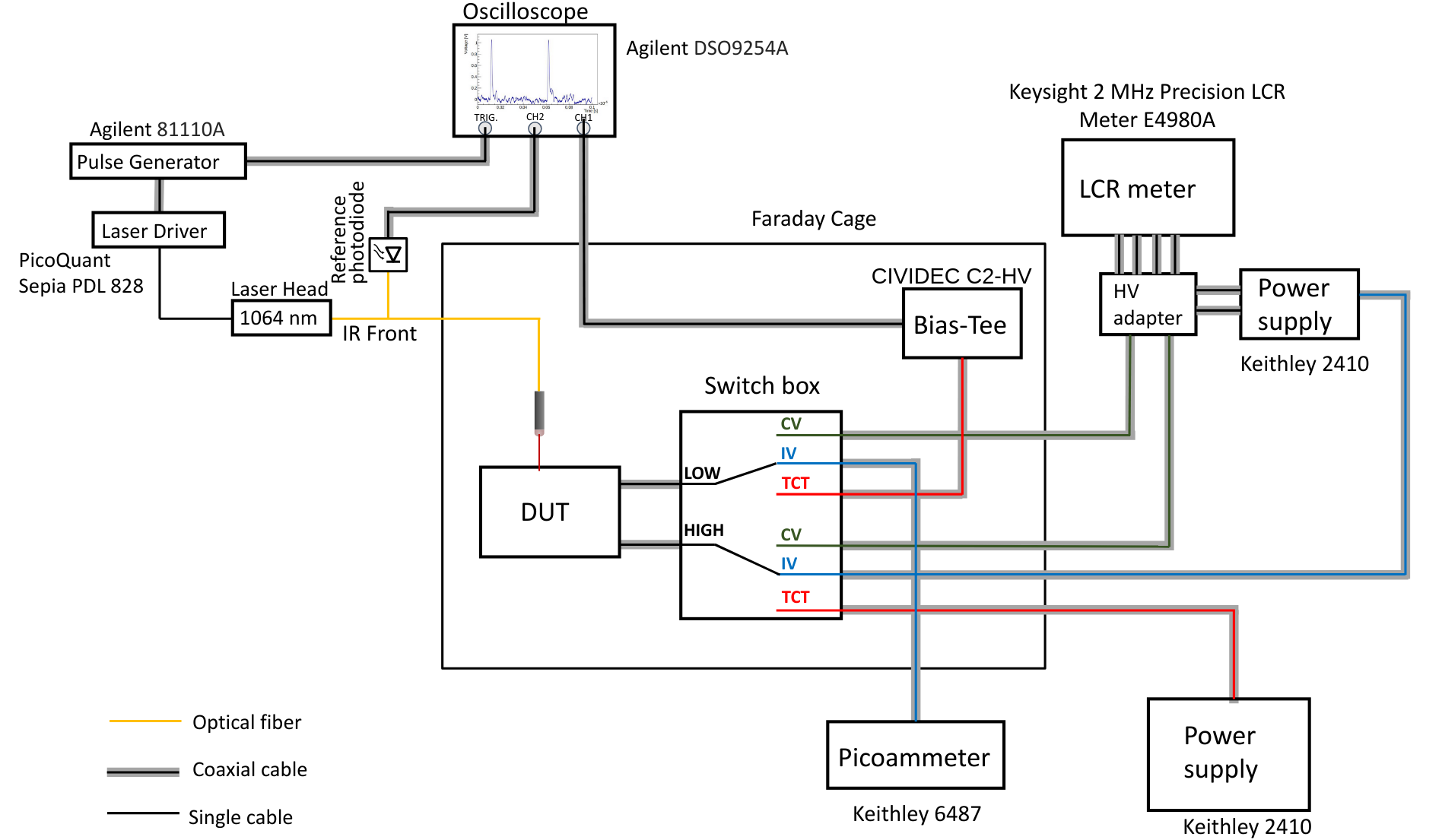}
    \caption{Diagram of the TCT+ setup~\cite{Kaluzinska:2815709}.}
    \label{fig:diagram_setup}
\end{figure}

The diagram of the TCT+ setup is shown in figure~\ref{fig:diagram_setup}.
The schematic of the Particulars setup is equivalent to the TCT+ setup schematic, shown in figure~\ref{fig:diagram_setup}, except for the brand of the amplifier, laser head, laser driver and pulse generator (all supplied by Particulars).
Both setups offer the capability for electrical characterization in addition to TCT measurements, enabling the measurement of leakage current (IV) and capacitance (CV) as a function of the reverse bias voltage.
More details about IV and CV measurements can be found in~\cite{Kieseler_2023}.  

\begin{figure}
    \centering
    \begin{subfigure}{0.44\textwidth}
        \includegraphics[width=0.85\textwidth]{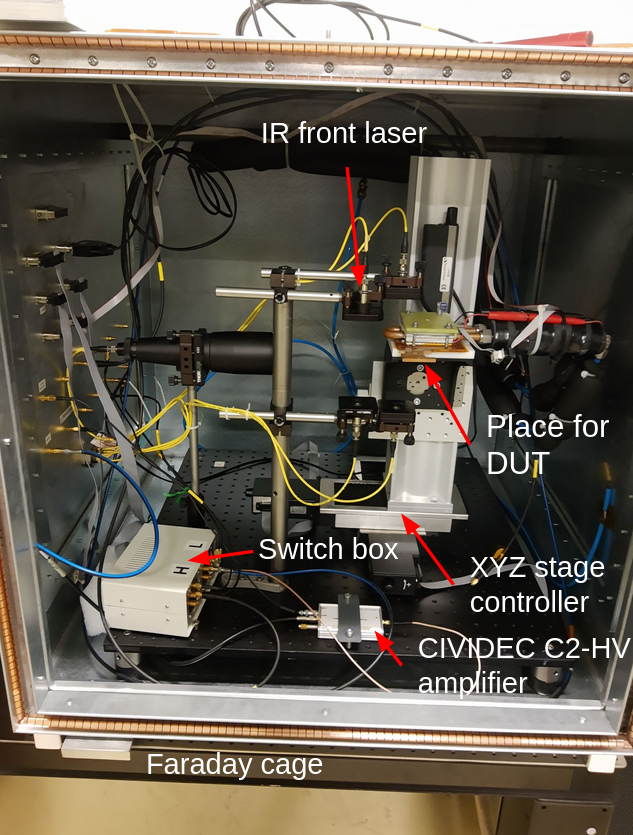}
        \caption{TCT+ setup.}
        \label{fig:setup_TCT+}
    \end{subfigure}
    \begin{subfigure}{0.54\textwidth}
        \includegraphics[width=\textwidth]{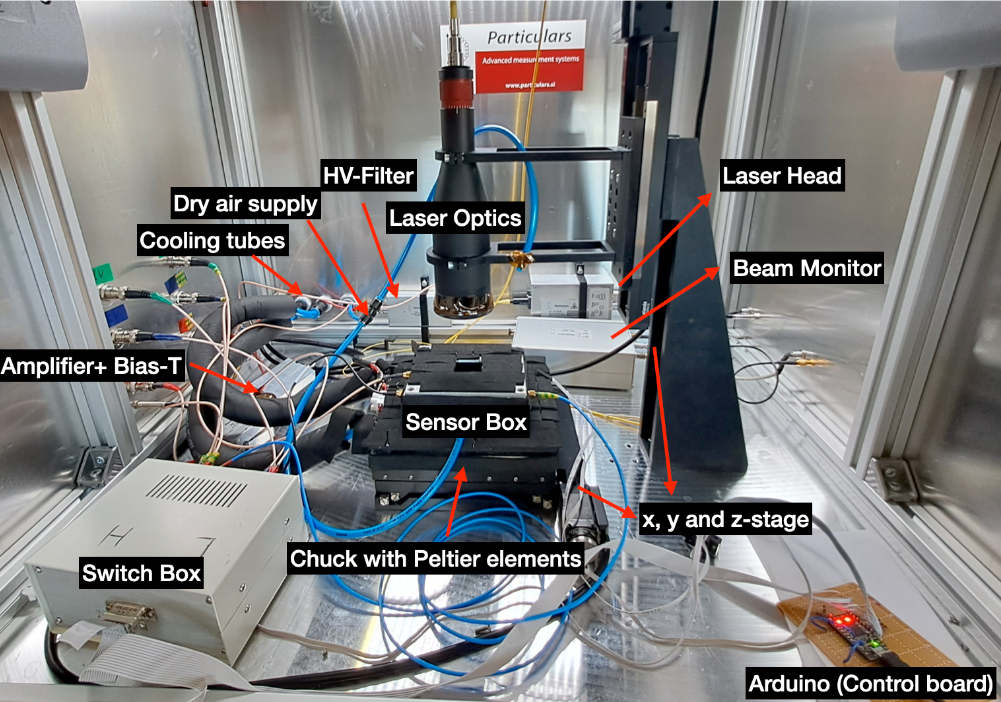}
        \caption{Particulars setup.}
        \label{fig:setup_Particulars}
    \end{subfigure}
    \caption{Two TCT setups used to obtain data reported in this publication, not showing the instruments located outside of the Faraday cages. Reproduced from \cite{Kaluzinska:2815709}.}
    \label{fig:setups}
\end{figure}

Both the TCT+ (figure~\ref{fig:setup_TCT+}) and Particulars (figure~\ref{fig:setup_Particulars}) setups serve the same purpose but differ in specific details.
In both setups, the DUT is mounted on a PCB attached to a copper holder or plate, which is cooled by Peltier elements connected to a chiller.
To control positioning, the TCT+ setup uses an XYZ stage that moves the copper holder while keeping the laser fixed, whereas the Particulars setup uses an XY stage to position the DUT and a separate Z stage to adjust the laser focus.
Both setups use a broadband current amplifier, the TCT+ setup is equipped with a CIVIDEC C2-HV amplifier with 43 dB gain and integrated bias-T~\cite{cividec}, while the Particulars setup uses a wide-band amplifier with 53 dB gain by Particulars company~\cite{particulars_amp}.
A switch box in both setups allows for changing between TCT, IV, and CV measurements.
To prevent condensation and frost, both setups are housed in a Faraday cage (as shown in figure~\ref{fig:setups}) and flushed with dry air.
Additionally, in the Particulars setup, dry air also flushes the enclosed sensor box, where the DUT is placed.
Both setups include a reference photodiode to monitor laser intensity, but the beam splitting differs.
In the TCT+ setup, a fused coupler from OZ Optics~\cite{beam_splitter_TCT} splits the beam, directing 90\% of the laser light to the DUT and 10\% to the photodiode.
In the Particulars setup, a fiber splitter inside the beam monitor evenly divides the intensity, sending 50\% to the DUT and 50\% to the photodiode.
The Particulars setup also contains an HV filter (to remove the noise in the HV line coming from the power supply) and external bias-T, which are both in the TCT+ circuit between the power supply and the sensor.
These setups have been cross checked in a dedicated study to assure there is no influence in the results.
\section{Data analysis procedure}
\label{sec:analysis}

\begin{figure}[h]
    \centering
    \begin{subfigure}[t]{0.5\textwidth}
        \includegraphics[width=\textwidth]{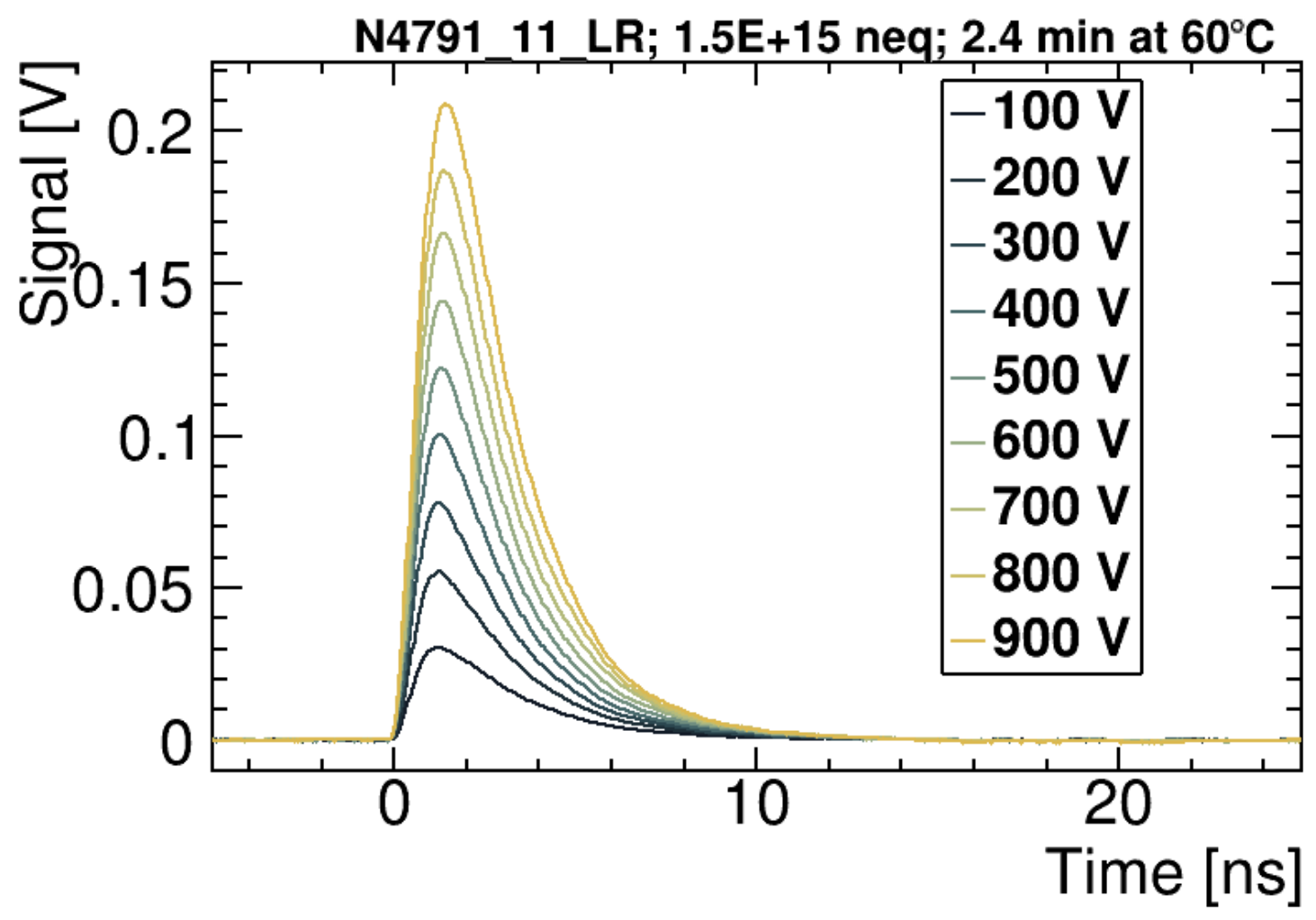}
        \caption{TCT waveforms for various bias voltages.}
        \label{fig:TCT_singal-example}
    \end{subfigure}
    \hfill
    \begin{subfigure}[t]{0.49\textwidth}
        \includegraphics[width=\textwidth]{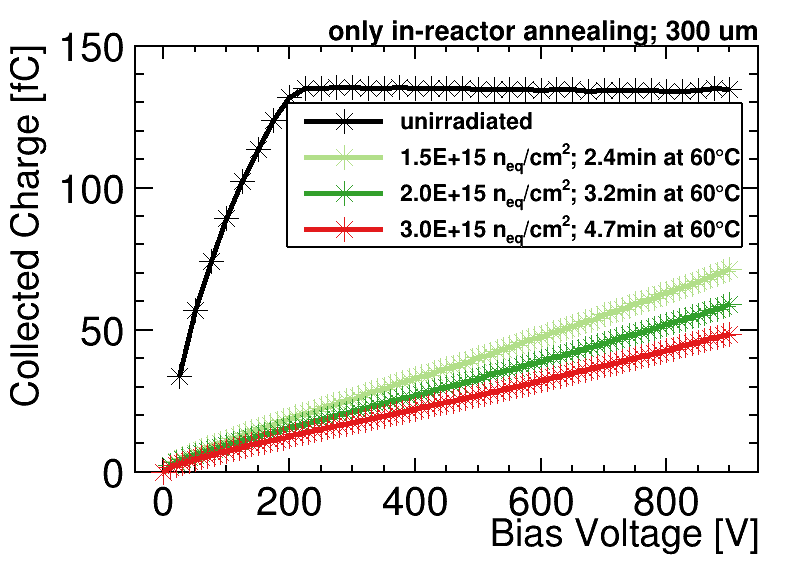}
        \caption{Integral of TCT waveforms building the collected charge value versus bias voltage.}
        \label{fig:CC_Vbias}
    \end{subfigure}
    \caption{Example of data processing on 300 \textmu m samples without additional annealing, obtained during the 2023 TCT+ campaign.}
    \label{fig:data_processing}
\end{figure}

\paragraph{Charge collection}
To determine the charge collection value, the TCT signal is measured and recorded.
The transient current pulse is converted into a voltage pulse through the input impedance (50 \ohm) of the amplifier.
Figure~\ref{fig:TCT_singal-example} displays an example of TCT signals obtained during the 2023 TCT+ campaign from an irradiated diode at various bias voltages up to the chosen measurement limit of 900 V.
This limit was chosen to reduce the risk of destructive discharges.
The signal's amplitude correlates with the applied bias voltage on the DUT.
In both TCT+ setup campaigns, the waveform measured at a particular bias voltage utilizes an average waveform, representing an average of 1000 samples.
For the 2023 Particulars campaign, 300 averaged waveforms are captured for each bias voltage, with a single waveform averaging 50 samples.

The collected charge value at a specific bias voltage is calculated as the integral of the TCT signal waveform (for a time window at 0 ns to 25 ns) divided by the amplifier gain.
An example of the charge collection vs.\ bias voltage dependence is depicted in figure~\ref{fig:CC_Vbias}.
After this initial data processing, the charge collection at a selected bias voltage can be analysed as a function of fluence or annealing time.

\paragraph{Saturation voltage}
In figure~\ref{fig:CC_Vbias}, it is clearly visible that the charge collection for unirradiated samples saturates after reaching a certain voltage.
This point is identified as the depletion voltage, which is determined using the same method as in CV measurements~\cite{Kieseler_2023}: by fitting the rising part of the curve and the plateau region, with the intersection point of these two fits representing the extracted depletion voltage.
However, for irradiated sensors, extracting the depletion voltage is significantly more challenging.
It is noteworthy that we distinguish "saturation voltage" from "depletion voltage".
This distinction arises from our observation that the concept of depletion voltage no longer holds true for irradiated devices since for example its value depends on the measurement frequency (for CV measurement) and the temperature~\cite{Kieseler_2023, CAMPBELL2002402}.

\begin{figure}
    \centering
    \includegraphics[width=0.59\textwidth]{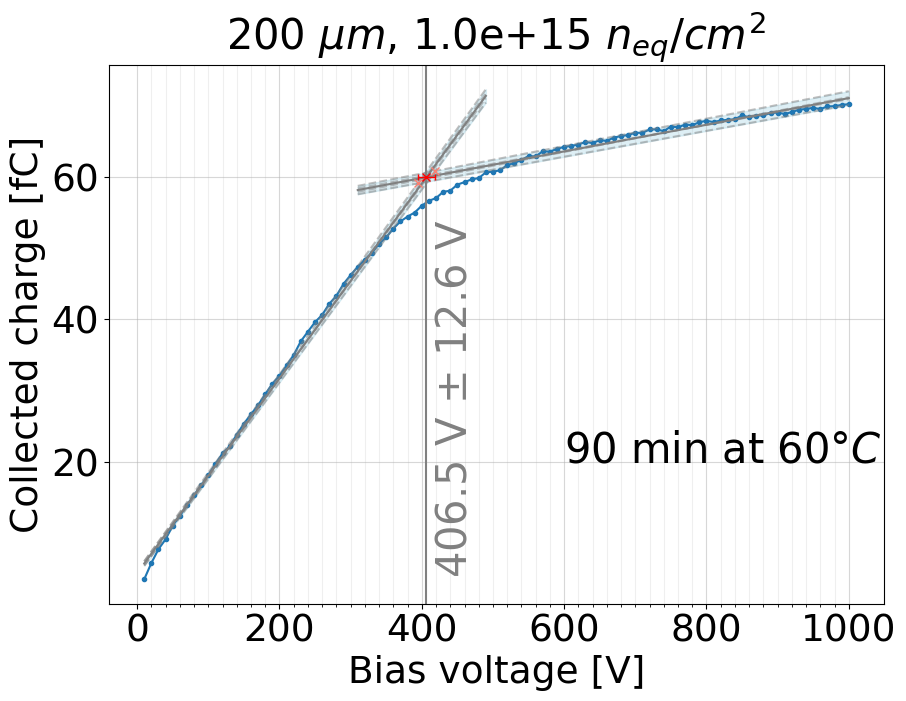}
    \caption{Saturation voltage extraction example. Data obtained during 2021 TCT+ campaign.}
    \label{fig:Vdep_extr_example}
\end{figure}

For the low fluences and thin sensors covered in this study, where the saturation voltage lies well below the measurement limit of 900 V, extracting the saturation voltage from charge collection measurements becomes feasible.
An example is shown in figure~\ref{fig:Vdep_extr_example}.
Unlike in unirradiated sensors, the plateau region is absent in irradiated sensors due to the occurrence of charge trapping, of which the probability decreases with increasing bias voltage and hence the electric field.
This results in an increase in charge beyond the saturation voltage.
Each part is fitted with a linear fit, and the intersection point is identified as the extracted saturation voltage.
If the second rising part is not reached within the measured voltage range or the two slopes become too similar, it becomes impossible to extract the saturation voltage (as visible in figure~\ref{fig:CC_Vbias} for the irradiated samples).
\section{Systematic uncertainties}
\label{sec:uncertainties}

Systematic uncertainties of charge collection were determined by considering five factors present during the campaigns, with two being the most significant: the laser and the temperature stability.
Long-term measurements of the laser intensity in the TCT+ setup, recorded by the reference photodiode (ref.\ section~\ref{sec:setups}), revealed a small instability over time.
To account for this instability, all the charge collection results obtained using the TCT+ setup in this study were calibrated by normalizing the laser intensity during the measurements to the intensity recorded during a reference measurement with a 300~\textmu m unirradiated diode.
After normalization, the uncertainty of the laser intensity was determined to be 2.5\%

Next, we considered the uncertainty in the sensor's temperature during measurements, which affects the laser absorption in the silicon.
In the TCT+ setup, the temperature is controlled within $\rm \pm 1\degree C$.
The 25e PCB layout used for the 2021 TCT+ campaign was larger than the copper holder in the setup, leading to temperature inhomogeneity across the PCB.
Additionally, it was observed, using an infra-red camera, that the SMA connectors were warmer than the middle part of the PCB, where the diode was located.
The PT1000 sensor's proximity to the SMA connector (ref.\ figure~\ref{fig:PCB25}) resulted in higher temperature readings.
This suggests that the diode could have been cooled more than intended. \\
The laser band-to-band absorption ($\alpha_{bb}$) depends on the temperature as follows~\cite{10.1063/1.4923379, GREEN20081305, KGSvantesson_1979}:
\begin{equation}
    \alpha_{bb}(T)=\alpha_{bb}(T_0)(T/T_0)^{c_{T}T_0}
\end{equation}
where $T_{0}$ is the nominal temperature for the absorption coefficient data (-20\degree C in our case) and $c_{T}$ is the temperature coefficient.
Based on the infra-red images of the setup, we assume that the sensor temperature during the 2021 TCT+ campaign was between -24\degree C and -19\degree C.
This translates to an uncertainty in charge collection due to temperature change of +6.1\% and -1.6\%.
For the 2023 TCT+ campaign, smaller PCBs were used (ref.\ figure~\ref{fig:PCB15}) which allows improved temperature control.
However, since the DUTs were already mounted to the PCBs, it was not possible to place a PT1000 directly onto them without the risk of damaging the DUT in the soldering process and uncontrolled annealing.
Instead, a PT1000 was attached to an additional copper sheet placed between the copper holder of the TCT+ setup and the PCB.
This increased the uncertainty of  the temperature measurement.
For this campaign, the DUT's temperature is assumed to be between -25\degree C and -18\degree C, resulting in an uncertainty in the charge collection due to temperature of +7.6\% and -3.2\%. 

There are two additional charge collection uncertainties considered for the TCT+ setup.
As the setup, during both campaigns, was also used by other users, it required everyday calibration.
This was achieved by measuring the charge collection of the 300 \textmu m unirradiated diode (as a reference) at +20\degree C at the start of each measurement day and adjusting the laser intensity to maintain consistent charge collection values.
The variations in charge collection resulted in a $\pm$0.8\% uncertainty for 2021 TCT+ campaign and a $\pm$0.6\% uncertainty for 2023 TCT+ campaign.
The second factor concerns the replacement of the amplifier in the setup on one occasion during 2021 TCT+ campaign and two occasions during 2023 TCT+ campaign, due to their failures.
After each amplifier replacement, the laser intensity was recalibrated in the same manner as during the daily calibration.
The variations in charge collection resulting from the amplifier replacements introduced a $\pm$0.2\% uncertainty for 2021 TCT+ campaign and $\pm$0.6\% uncertainty for 2023 TCT+ campaign. 

All uncertainties were combined in quadrature, resulting in a total charge collection uncertainty of +6.6\% and -3.1\% for the 2021 TCT+ campaign and +8.0\% and -4.2\% for the 2023 TCT+ campaign.

The Particulars setup was exclusively used for this measurement campaign, with no other users.
A temperature  uncertainty of the smaller PCB has to be considered ($\pm$1\degree C), as well as an uncertainty regarding the laser stability.
To monitor this, a reference 300 \textmu m unirradiated diode was measured daily.
It was found that environmental temperature variations slightly influence the laser intensity, as well as the temperature stability of the DUT.
To account for this, an overall charge collection uncertainty of 5.0\% was estimated.

The last contributing factor to the charge collection uncertainty is the fit uncertainty of the Gaussian fit applied to the charge collection distribution of the 300 recorded events, of which the most probable value defines the CC for each measured voltage.

\begin{table}
    \centering
    \begin{tabular}{c|c|c|c|}
         & 2021 TCT+ & 2023 TCT+ & 2023 Particulars\\
         \hline
       Laser instability  & $\pm$2.5\% & $\pm$2.5\% & * \\
       \hline
       Laser absorption (due to temperature)  & +6.1\% -1.6\% &  +7.6\% -3.2\% &$\pm$1.6\% \\
       \hline
       Everyday variations in charge collection  & $\pm$0.8\% & $\pm$0.6\% & *\\
       \hline
       Amplifier replacement  & $\pm$0.2\% & $\pm$0.6\% & --\\
       \hline
       Total  & +6.6\% - 3.1\% & +8.0\% -4.2\% & $\pm$5.0\%\\
    \end{tabular}
    \caption{Summary of all systematic uncertainties in charge collection, * -- for 2023 Particulars campaign the laser instability and everyday variations in CC are not distinguishable.}
    \label{tab:uncertainty_CC}
\end{table}

To assess the uncertainty in annealing time, we first account for the uncertainty in the estimated initial 60\degree C offset, which is derived by considering the full impact of the reactor temperature uncertainty.
These effects become more pronounced for longer irradiation times.
Considering the high total annealing times - up to 500 min - done in this campaign, the impact of the uncertainty of the in-reactor annealing becomes negligible.
Next, for the individual annealing steps, we assign a 0.5\degree C uncertainty in the temperature measurement, based on the temperature recordings.
The primary factor contributing to this uncertainty is the potential for small temperature variations between different PCBs.
The uncertainty in the time measurement itself is negligible.
All systematic uncertainties in charge collection are summarized in the table~\ref{tab:uncertainty_CC}.

The uncertainty on the fluence is consistently 10\% for all fluences.

In this paper, the extracted saturation voltages from both CV and CC measurements are presented as well.
The uncertainty associated with the saturation voltage from CV measurements is well described in~\cite{Kieseler_2023}.
To estimate the uncertainty on the extracted saturation voltage from CC measurements, an automated fitting method was employed (as shown in figure~\ref{fig:Vdep_extr_example}).
Since the data around the crossing point is not strictly linear, the uncertainty was assessed by varying the fit range by ±10\%, repeating the fit, and re-evaluating the crossing point.
The maximum upward and downward deviations from the original value were taken as the uncertainty range.
The contribution from the linear fit itself is negligible and thus not considered separately~\cite{Muehlnikel:2920469}.
\section{Results}
\label{sec:results}

\begin{figure}[ht]
    \centering
    \begin{subfigure}{0.5\textwidth}
        \includegraphics[width=\textwidth]{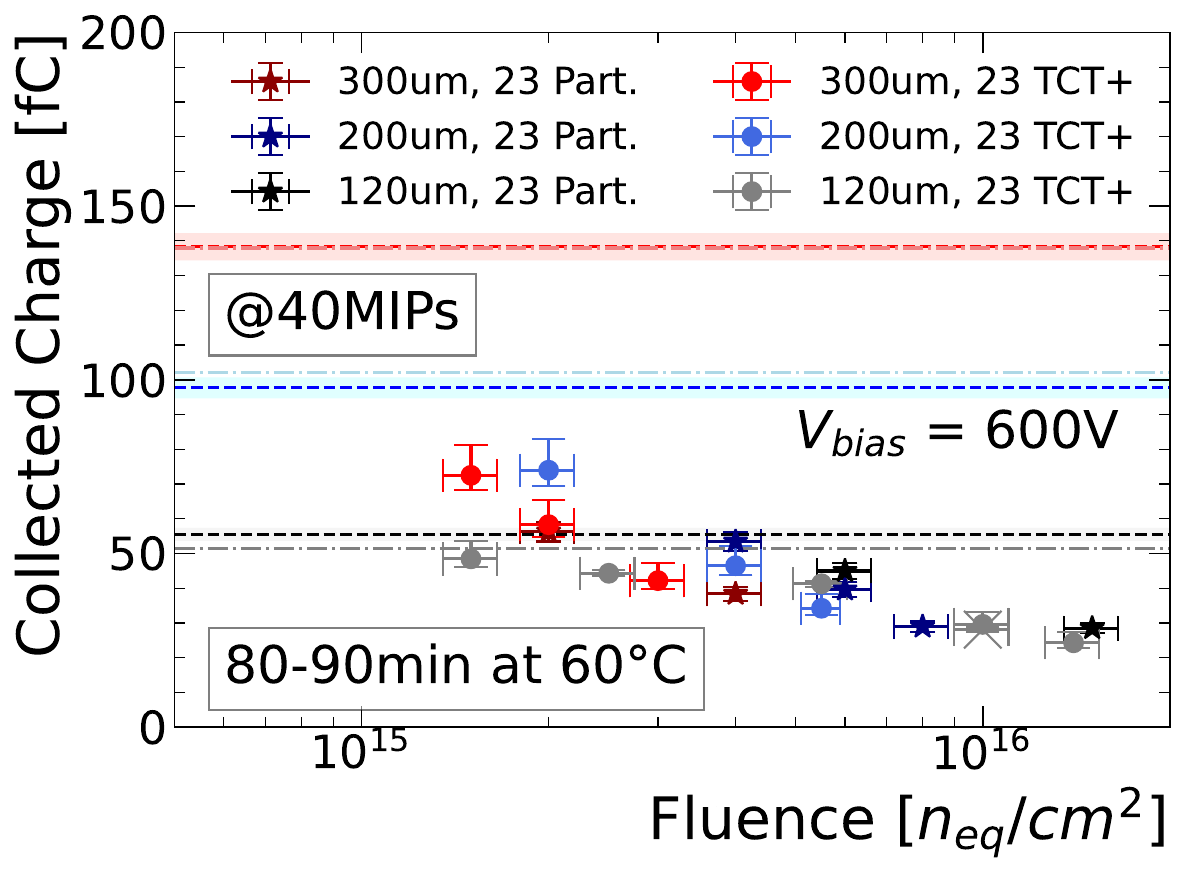}
        \phantomsubcaption
        \label{fig:CC_90min}
    \end{subfigure}
    \begin{subfigure}{0.49\textwidth}
          \includegraphics[width=\textwidth]{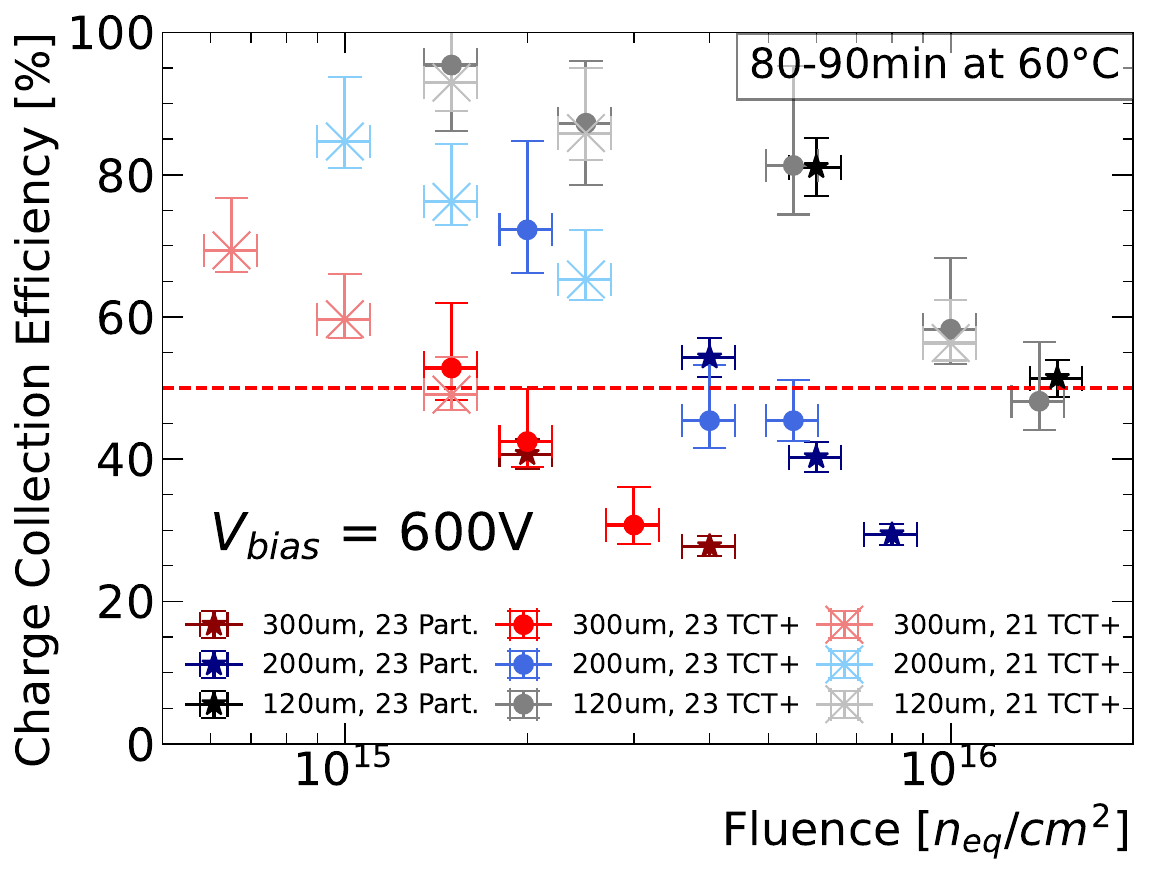}
        \phantomsubcaption
        \label{fig:CCE_90min}
    \end{subfigure}
    \vspace*{-6mm}
    \caption{Charge collection (a) and charge collection efficiency (n) as function of the fluence for all the campaigns.}
    \label{fig:Plots_90min_all}
\end{figure}

The data collected throughout different campaigns are used to characterize the behaviour of the charge collection (ref.\ section~\ref{sec:analysis}) over a broad range of fluence.
Since only an IR laser is used, the individual trapping times are not accessible within this measurement campaign~\cite{KRAMBERGER2002645, KRAMBERGER2007762, Adam_2016}.
Figure~\ref{fig:Plots_90min_all} depicts the charge collection behaviour at a bias voltage of 600 V after approximately 90-minute annealing at 60\degree C for all three sensor thicknesses.
In figure~\ref{fig:CC_90min}, only the charge collection data from the two 2023 campaigns are presented, while the results from the 2021 TCT+ campaign are omitted since a different in laser intensity was used (33 MIPs equivalent vs.\ 40).
The horizontal dotted lines represent the charge collection values for unirradiated sensors, with red indicating 300~\textmu m, blue representing 200~\textmu m, and black representing 120~\textmu m.
Comparing the absolute collected charge, at high fluences like $\rm2.0\cdot~10^{15} n_{eq}/cm^{2}$ or $\rm4.0\cdot~10^{15} n_{eq}/cm^{2}$, the 200~\textmu m sensors still collect roughly 20\% more charge than the 300~\textmu m sensors.
This can be attributed to the fact that a larger region of the detector is depleted for thinner sensors at a fixed voltage.
A decrease in collected charge with increasing fluence is consistently observed across all three thicknesses.
The collected charge values demonstrate agreement within the margins of uncertainty across different campaigns.

In figure~\ref{fig:CCE_90min}, the charge collection efficiency (CCE), defined as the charge collection of an irradiated sample divided by the charge collection of the non-irradiated reference sensors, across all three campaigns is depicted.
A red horizontal line at 50\% CCE is included for better visibility, providing an additional safety margin in case the LHC delivers a higher integrated luminosity.
Once more, a distinct trend emerges: as fluence increases, the charge collection efficiency decreases across all thickness variations.
The data collected from different campaigns and studies on earlier CE prototypes at lower fluences~\cite{Akchurin_2020} exhibit agreement within the margins of uncertainty.
At a fixed fluence, the 300~\textmu m samples demonstrate the lowest charge collection efficiency, whereas the 120~\textmu m samples exhibit the highest.
Since thinner sensors collect more charge (as shown in figure~\ref{fig:CC_90min}), especially at higher fluences, this indicates superior radiation hardness, making them suitable for deployment in the highest radiation regions of the CE.
Specifically, for 600 V, the 300~\textmu m samples achieve 50\% charge collection efficiency up to a fluence of $\rm1.5\cdot~10^{15} n_{eq}/cm^{2}$, the 200~\textmu m samples up to $\rm4.0\cdot10^{15} n_{eq}/cm^{2}$, and the 120~\textmu m samples up to $\rm1.5\cdot10^{16} n_{eq}/cm^{2}$.
At a fluence of $\rm1.5\cdot10^{15} n_{eq}/cm^{2}$, the 120~\textmu m samples still have nearly 100\% efficiency.

\begin{figure}[ht]
	\centering
	\begin{subfigure}{0.49\textwidth}
    	\includegraphics[width=\textwidth]{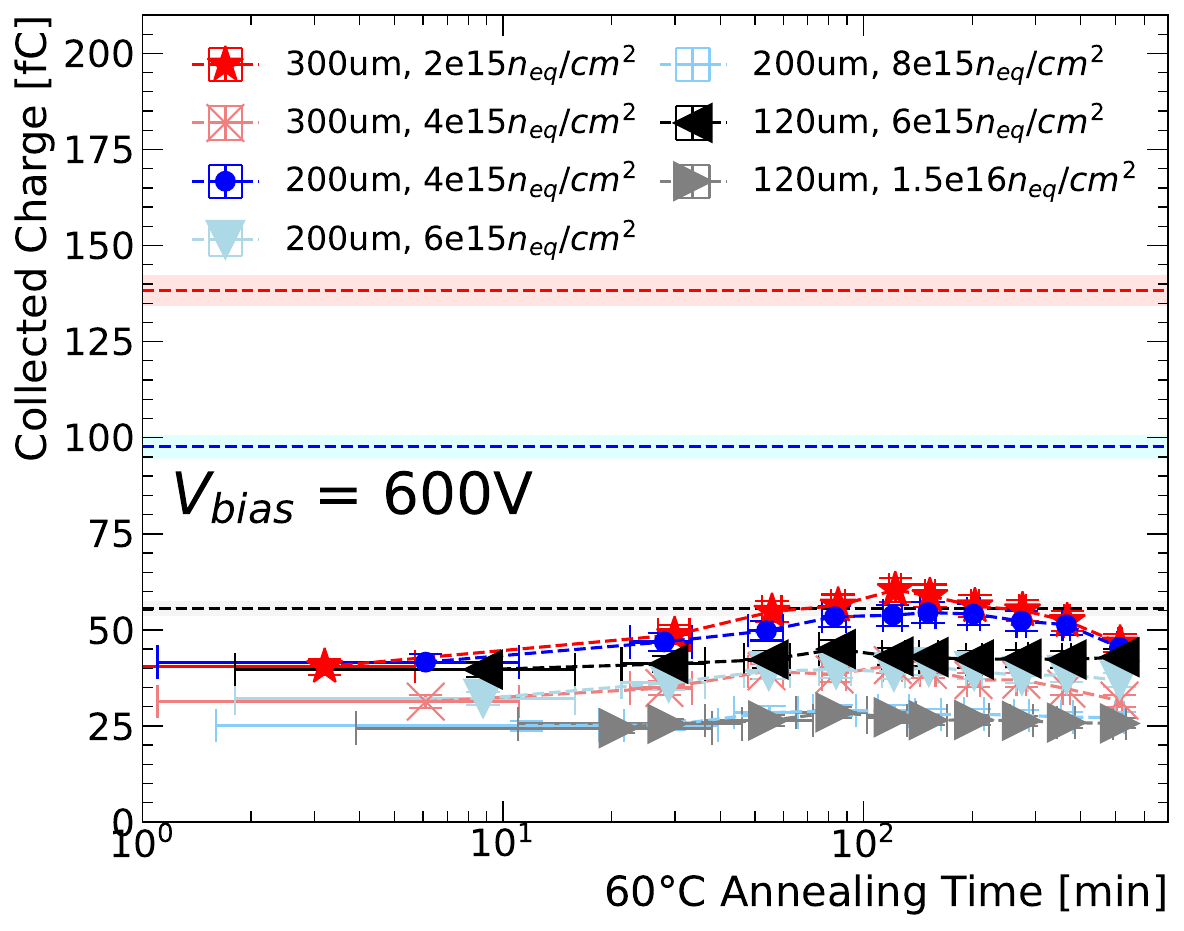}
    	\phantomsubcaption
    	\label{fig:CC_time}
	\end{subfigure}
	\hfill
	\begin{subfigure}{0.49\textwidth}
    	\includegraphics[width=\textwidth]{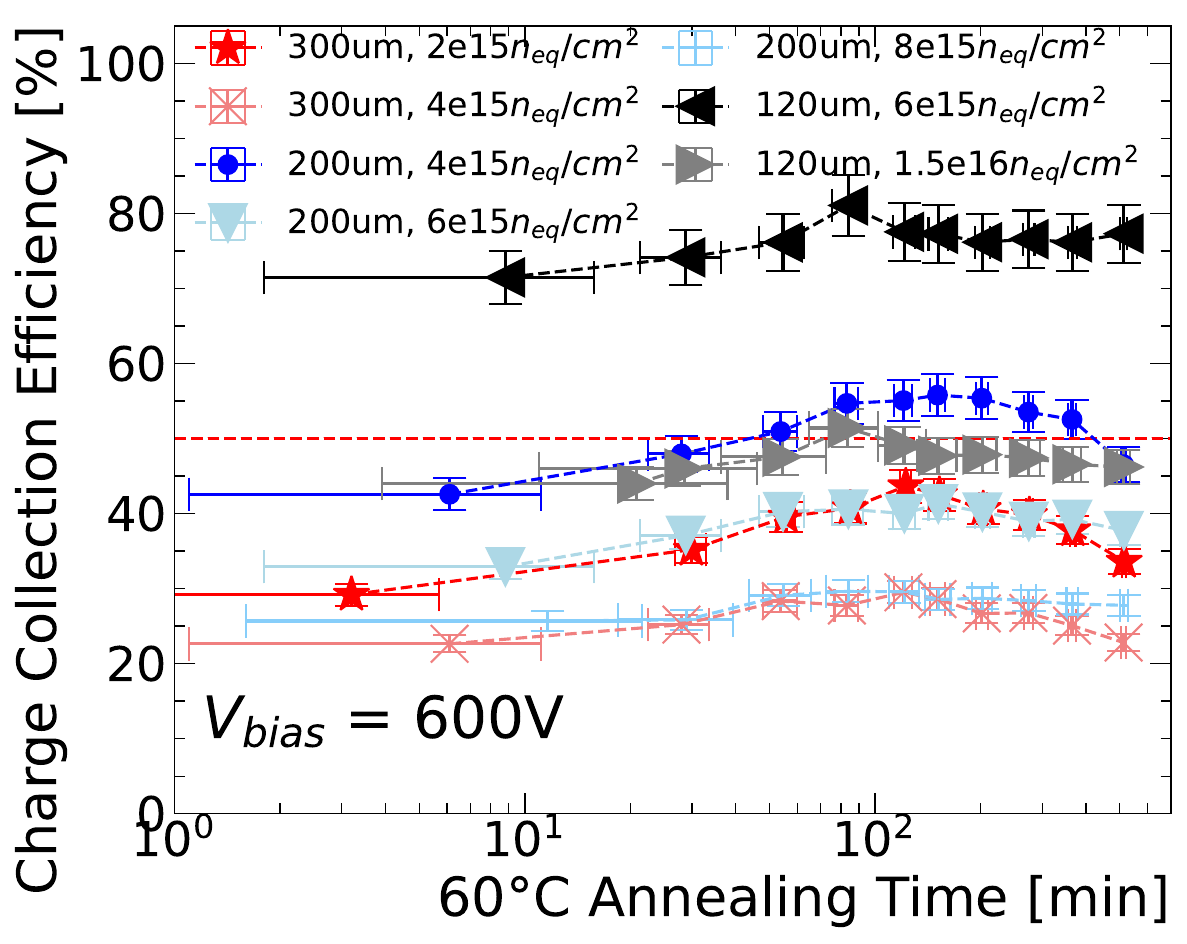}
    	\phantomsubcaption
    	\label{fig:CCE_time}
	\end{subfigure}
 	\vspace*{-6mm}
	\caption{Charge collection (a) and charge collection efficiency (b) as function of the annealing time. Data collected during 2023 Particulars campaign.}
	\label{fig:CC_time_all}
\end{figure}

\begin{figure}[h]
	\centering
	\includegraphics[width=0.6\textwidth]{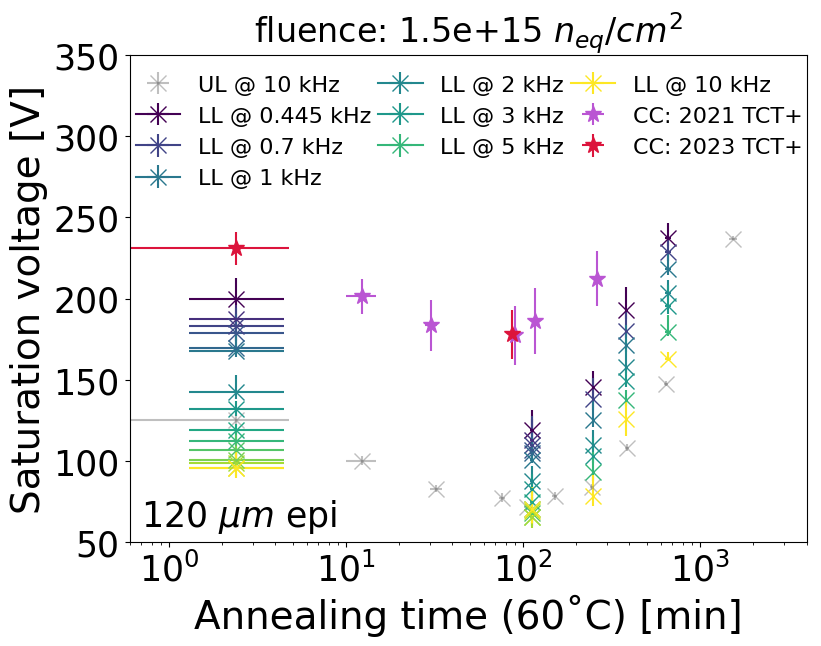}
	\caption{Saturation (former "depletion") voltage as a function of the annealing time at 60\degree C. Data extracted from CV measurements at several frequencies presented in~\cite{Kieseler_2023} and charge collection (CC) measurements (2021 TCT+ and 2023 TCT+ campaigns). 'UL' refers to the upper left, and 'LL' to the lower left halfmoon of the wafer (ref.\ fig.\ \ref{fig:HGCAL_full_wafer}).}
	\label{fig:Vdep_120_1p5e15}
\end{figure}

For the samples irradiated in 2023 and characterized in the Particulars setup the annealing was performed in 9 steps with TCT measurements between the steps.
A visible increase in collected charge (figure~\ref{fig:CC_time}) and efficiency of collected charge (figure~\ref{fig:CCE_time}) is observed for all samples during the time range dominated by beneficial annealing.
The maximum is reached at around 120 minutes for float zone samples and around 85 minutes for epitaxial samples followed by a decrease into time range dominated by reverse annealing.
These findings align with previous CC studies on p-type sensors~\cite{LDiehl_ptype,Adam_2020, Casse:6164289} and with the corresponding findings from CV measurements~\cite{Kieseler_2023}.

\begin{figure}[b]
	\centering
	\includegraphics[width=0.64\textwidth]{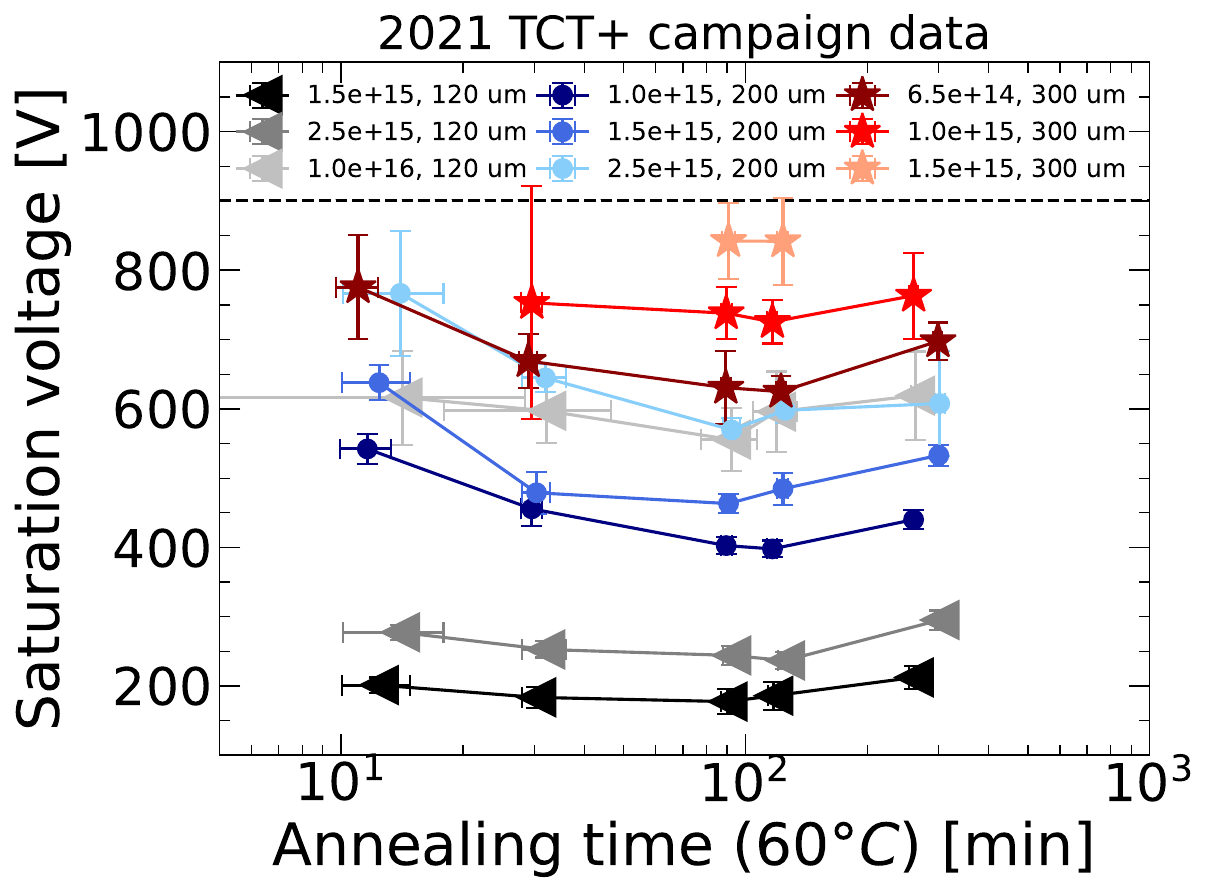}
	\caption{Saturation voltage as a function of annealing time at 60\degree C extracted from charge collection data from TCT 2021+ campaign. The measurement limit of 900 V is highlighted as black dashed line.}
	\label{fig:Vsat_anntime_all_2021}
\end{figure}

In previous studies~\cite{Kieseler_2023}, the saturation voltage (ref.\ section~\ref{sec:analysis}) extracted from capacitance versus bias voltage (CV) measurements, as a function of annealing time, was used as an input to the Hamburg model.
However, the saturation voltage obtained from CV measurements is well known for its sensitivity to both frequency and temperature~\cite{CAMPBELL2002402, Kieseler_2023}.
To mitigate this dependency, we use charge collection as an independent alternative approach.
Figure~\ref{fig:Vdep_120_1p5e15} illustrates a comparison between the depletion voltage extracted from CV and charge collection measurements, using a sample with the following properties: 120 \textmu m thickness, irradiated to $\rm1.5\cdot10^{15} n_{eq}/cm^{2}$.
The~charge collection was measured during two campaigns: 2021 TCT+ and 2023 TCT+.

It is evident that, as the CV frequency increases, the extracted saturation voltage decreases.
Moreover, the saturation voltage extracted from charge collection is higher compared to those obtained from CV with the lowest CV frequency being 445 Hz.
This poses a challenge for thicker and highly irradiated sensors, where the saturation voltage is correspondingly higher.
If the saturation voltage exceeds 900~V, it becomes infeasible to measure it using the CC method with the mentioned bias voltage limitation, whereas for CV it can be tuned down by frequency choice.
However, the extraction of the saturation voltage using the CC method is unaffected by measurement settings such as the frequency choice of the CV measurement, which is considered an advantage.
For the temperature range -30\degree C to -15\degree C the saturation voltage for irradiated samples extracted from charge collection measurements stays constant within 0.1\%, which was tested with dedicated measurement within this study (figure not shown).

The extraction of the saturation voltage from charge collection data is feasible for samples where the saturation voltage is well below the measurement limit of 900 V which is true for the thin samples and low fluence 300~\textmu m sample for the TCT+ 2021 campaign.
The extracted saturation voltages are presented in figure~\ref{fig:Vsat_anntime_all_2021}.
The minimum saturation voltage is observed between 90 and 120 minutes, consistent with results from CV measurements~\cite{Kieseler_2023}.
The results obtained from the measurement of the saturation voltage as well as the leakage current for different annealing temperatures can be used as inputs to the Hamburg model~\cite{Moll:2018fol}.
Even though not all assumptions are made for that model, it still provides a good parameterization to extract the temperature dependence and helps to predict the evolution of parameters like charge collection efficiency and leakage current -- which is important for noise and power consumption -- throughout the operational lifetime of the experiment.
However, such study benefits from extending the annealing campaign results to fluences which are lower than the ones reported here.
Hence it is beyond the scope of this paper, but considered in ongoing campaigns.
\section{Conclusions}
\label{sec:discussion}

The charge collection annealing behaviour of silicon diodes from 8-inch p-type wafers from the CE prototypes and pre-series phases is presented for a large fluence range.
The combined results from three charge collection campaigns confirm that the bulk material of diodes from different CE manufacturing phases exhibit consistent behaviour, as expected. 

The results obtained in this study provide valuable input for optimizing the CE detector layout.
Although thick sensors have a lower radiation tolerance, they benefit from an initially higher signal-to-noise ratio.
Based on the results hereby reported, on the measurements of the leakage current~\cite{Kieseler_2023} and on the expected performance of the front-end readout chip to be used in CE~\cite{Bouyjou_2022}, one can extend the usage of thicker sensors in higher fluence regions.
The previously envisaged layout for the CE planned to use 300~\textmu m sensors only in regions that do not expect higher fluences than $\rm5.0\cdot10^{14} n_{eq}/cm^{2}$ after an integrated luminosity of 3.0 $\rm ab^{-1}$, while the 200~\textmu m sensors were planned to cover regions expecting fluences up to $\rm2.5\cdot10^{15} n_{eq}/cm^{2}$~\cite{HGCAL-TDR}. After this dedicated high-fluence study, the layout was optimized to use 300~\textmu m sensors in regions up to $\rm1.7\cdot10^{15} n_{eq}/cm^{2}$ and 200~\textmu m sensors up to $\rm5.0\cdot10^{15} n_{eq}/cm^{2}$, so that the detector benefits from initially larger signals and lower production costs. These fluence limits were calculated based on the measured performance taking a safety margin regarding maximum integrated luminosity into account.

The cooling scheme used for the CE will ensure the operation at -35\degree C and never exceeding (-30\degree C).
During the shutdown periods it will be warmed up to approximately 0\degree C.
This study supports the case that the silicon sensors will not enter the reverse annealing-dominated region significantly during the CE's operational lifetime.
Our findings show that the charge collection behaviour of the novel 8-inch material with respect to annealing time reaches its maximum increase at around 120 minutes for float zone material and around 85 minutes for epitaxial material at 60\degree C.
These results align with CV measurements and other studies on 6-inch p-type sensors, indicating that the CE 8-inch sensors will not experience significant reverse annealing until the end of HL-LHC if we apply the Hamburg model with the parameters values derived in~\cite{Moll:1999kv}, which allows extrapolating down to low temperature.

The presented dataset is currently being extended with an annealing campaign at additional temperatures (6.5\degree C, 20\degree C, 30\degree C, 40\degree C, 60\degree C) and annealing steps.
Due to the slower annealing at the lower temperatures, results are expected in 2025/2026.
This approach will enable to perform the Hamburg model fit at various annealing temperatures and extract for the first time scaling factors between the extracted parameters for the 8-inch p-type material.
Consequently, it will be possible to predict the annealing behaviour of CE silicon sensors at 0\degree C with better accuracy as using the current Hamburg model and establish appropriate CE operating conditions.
\acknowledgments

We thank the CERN EP-DT SSD group for providing the TCT+ setup to perform the measurements described in this publication, in particular Michael Moll, Ruddy Costanzi, Esteban Curras Rivera, Marcos Fernandez Garcia and Moritz Wiehe.
Futhermore we thank Dana Groner and Ruddy Costanzi for help in upgrading the Particulars setup.
Moreover, we thank the CERN EP R\&D programme for funding the Particulars setup and the CMS HGCAL silicon sensor group for the discussions about the results presented here.
We are grateful for feedback and suggestions from Rachel Yohay, Ronald Lipton, Pedro Silva and Philippe Bloch.
This work has been sponsored by the Wolfgang Gentner Programme of the German Federal Ministry of Education and Research (grant no. 13E18CHA).
This work has been supported by the Alexander-von-Humboldt-Stiftung.

\bibliographystyle{JHEP}
\bibliography{main.bib}

\end{document}